\documentclass{article}[12pt]
\usepackage{epsf}
\usepackage{psfig}
\usepackage[dvips]{color}
\def\gtrsim{\mathrel{\hbox{\rlap{\hbox{\lower4pt\hbox{$\sim$}}}\hbox{$>$}}}}
\def\lesssim{\mathrel{\hbox{\rlap{\hbox{\lower4pt\hbox{$\sim$}}}\hbox{$<$}}}}
\begin{document}
\begin{flushleft}
{\bf Origin of the Hard X-ray Emission from the Galactic Plane}
\end{flushleft}

\medskip

\begin{flushleft}
Ken Ebisawa\footnote{ code 662, NASA/GSFC, Greenbelt, Maryland 20771, USA, and Universities Space Research
Association}$^*$, Yoshitomo Maeda\footnote{ Department of Astronomy \& Astrophysics, Pennsylvania State University,
525 Davey Lab.,  University Park, Pennsylvania      16802, USA:
Present address; 
Institute of  Space and Astronautical Science, 3-1-1 Yoshinodai, Sagamihara, Kanagawa 229-8510, Japan}, 
Hidehiro Kaneda\footnote{Institute of  Space and Astronautical Science, 3-1-1 Yoshinodai, Sagamihara, Kanagawa 229-8510, Japan}, 
Shigeo Yamauchi\footnote{Faculty of Humanities and Social Sciences, Iwate University, 3-18-34 Ueda, Morioka, Iwate 020-8550, Japan\\
$^*$ To whom correspondence should be addressed. E-mail: ebisawa@gsfc.nasa.gov}
\end{flushleft}

\begin{flushleft}
{\bf 
The Galactic plane is a strong hard  x-ray  emitter 
and the emission forms a narrow continuous ridge ({\em 1-3}).
The currently known hard x-ray sources are far too few to explain
the ridge x-ray emission, and  the fundamental question as to
whether the ridge emission is ultimately resolved into numerous dimmer discrete
sources or truly diffuse emission has not yet been settled
({\em 4-9}).
In order to obtain a decisive answer, 
using the Chandra x-ray observatory, we have carried out the deepest
hard x-ray survey of a  Galactic plane region which is devoid of known
x-ray point sources.
We have detected at least 36 new hard x-ray point sources in addition to 
strong diffuse emission within a $17' \times
17'$ field of view.
The surface density of the point sources is comparable to that at
high Galactic latitudes after the effects of Galactic absorption are considered. 
Therefore, most of these point sources are probably  extragalactic,
presumably   active galaxies  seen through the Galactic disk.
The Galactic ridge hard x-ray emission is diffuse,
which indicates  omnipresence of the hot plasma along the Galactic plane
whose energy density is more than one order of magnitude higher than 
any other substances in the interstellar space.
}

\end{flushleft}

The Galactic ridge x-ray emission exhibits emission lines from highly 
ionized heavy elements  such as Si, S and Fe, hence it may be considered from 
optically thin hot plasmas with a temperature of several keV ({\em 3}).
If the  plasma distribution is diffuse in the Galactic disk,
the plasma temperature is higher than that can be bound by Galactic gravity, and 
its energy density, $\sim$ 10 eV/cm$^3$, 
is one or two orders of magnitude 
higher than those of other constitutes in the interstellar space, 
such as cosmic rays, Galactic magnetic fields, and ordinary
interstellar medium  ({\em 3, 8}).
Another hypothesis is that the  Galactic ridge x-ray emission is a superposition of numerous 
point sources ({\em 1, 4, 5, 6}). However, 
 no class of x-ray objects is known with such a high plasma temperature 
and a large number density to satisfy the uniform surface 
brightness of the ridge emission ({\em 8, 9}).

To  resolve  the origin of the Galactic ridge x-ray emission,
we observed a ``blank'' region of the Galactic plane,  
$(l,b) \approx (+28^\circ.45,-0^\circ.2)$,
where the Advanced Satellite for Cosmology and Astrophysics
(ASCA) could not  find any point sources ({\em 7, 8})
brighter than $\sim 2 \times 10^{-13}$ erg s$^{-1}$ cm$^{-2}$ (2 -- 10 keV).
We used the  Advanced CCD Imaging Spectrometer Imaging-array
(ACIS-I) on-board the Chandra Observatory, 
with unprecedented sensitivity and imaging quality  (Fig. 1).
The observation was carried out on  25 and 26 February, 2000,
for a  total exposure time of 90 ksec.
The pointing position was chosen because the direction is tangential to the 
Scutum arm  where  the  ridge x-ray emission  is  strong.

We were interested in hard x-ray emission, so
we searched  in the 3 -- 8 keV band 
to minimize the effects  of  Galactic absorption at lower energies 
and the intrinsic non-x-ray background in the higher energy band.
The ``wavdetect'' source finding 
program in the Chandra data analysis package
detected 53, 36 and 29  sources in the $17'\times17'$ ACIS-I field with 3, 4 and
4.5  $\sigma$ significance, respectively.  For each of  these sources,
we determined the energy flux in the 2 -- 10 keV band by
fitting the energy spectrum after accounting for the position dependent 
detector responses.
The 3, 4 and 4.5 $\sigma$ thresholds  roughly correspond
to energy fluxes of $\sim 3 \times10^{-15}$,
$\sim 4 \times10^{-15}$ and $\sim 5 \times10^{-15}$ ergs s$^{-1}$ cm$^{-2}$
in the 2 -- 10 keV band, respectively.  
%Our main result
%in this paper does not depend on the choice of these  threshold levels.
We found point sources with different fluxes
are randomly distributed over the field of view, which indicates that the
positional dependence of the source detection efficiency is negligible.
%To make the comparison with other
%observations easier, below we show the flux  converted to the 
%2 -- 10 keV band using this assumed spectrum.
We have also carried out a source search in the 0.5 -- 3 keV soft x-ray band,
and detected 106 sources with 3 $\sigma$ confidence.
There are only 17 sources which are detected both in the hard and soft band
over 3 $\sigma$ confidence,
which suggests that the populations  of the soft sources and
hard sources are  different.  We
 identified several soft x-ray sources, which are not detected in the hard band,
with objects in the  United States Naval Observatory A2.0
catalog and/or the Palomar Digital Deep Sky Survey; 
thus these sources are probably ordinary stars. On the other hand, 
none of the hard x-ray sources we observed have been identified.

Chandra data consist of not only x-ray events from point sources and diffuse 
emission, but also  non-x-ray background events.
Typical non-x-ray background rate has been calculated and
released by Chandra X-ray Center, based on observations
of source-free high Galactic latitude regions.
By subtracting the expected non-x-ray background rate, 
we have determined the total hard x-ray flux in our field of view
as  $\sim 1.1 \times 10^{-10}$ ergs s$^{-1}$ cm$^{-2}$ deg$^{-2}$ in 2 -- 10 keV,
which includes contributions from both diffuse emission and point sources.
On the other hand, integrated  hard x-ray flux of all the point 
sources above the flux 3 $\times 10^{-15}$ ergs s$^{-1}$ cm$^{-2}$ is 
$\sim 9.8 \times 10^{-12}$ ergs s$^{-1}$ cm$^{-2}$ deg$^{-2}$ (2 -- 10 keV), 
which is only $\sim$ 10 \% of the total 
observed x-ray flux in the field of view, and the rest is the diffuse emission
(Fig.\ 1 and 2).  

The point x-ray sources on the Galactic plane may comprise
extragalactic sources seen through the Galactic plane and 
Galactic sources.
Remarkably, the surface density of the hard x-ray sources on the Galactic plane
we have determined is consistent with 
that of the high Galactic latitude fields 
in a  similar flux range (Fig.\ 2).  
The x-ray fluxes of extragalactic sources are reduced on the Galactic plane
because they are absorbed  by  interstellar matter.
The Galactic HI column density in our Chandra pointing 
direction is $\sim 2 \times 10^{22}$ cm$^{-2}$ ({\em   12})
and that of the molecular hydrogen is $(1 - 2) \times 10^{22}$ cm$^{-2}$ ({\em  13}), both
measured from radio observations ({\em  12,  13}).
Therefore, the total  hydrogen column density through the
Galactic plane is $N_{H} = N_{HI} + 2 N_{H_{2}} = (4-6) \times
10^{22}$ cm$^{-2}$.
Even if we account for the $\sim$ 30 \% flux reduction of the extragalactic 
hard x-ray sources caused by the  interstellar absorption of $N_H
= 6\times 10^{22}$ cm$^{-2}$, 
the extragalactic $\log N - \log S$ curve is still consistent with 
the present Chandra
Galactic $\log N - \log S$ curve within 90 \% 
statistical uncertainty (Fig.\ 2).
Therefore, most of the point sources detected in our field must be
 extragalactic, presumably Active Galactic Nuclei, which  dominate the
cosmic x-ray background ({\em  10}).

From our observation,  the point source density 
on the Galactic plane in the flux range above $ 3\times10^{-15}$ ergs s$^{-1}$ cm$^{-2}$ (2 -- 10 keV)
is 660$\pm$160 sources/deg$^2$ (90 \% error), among which
$\sim$ 560 sources/deg$^2$ is considered to be the extragalactic sources
 ({\em  10}), where we took  account of the $\sim 30 $ \% flux reduction
due to the Galactic absorption.  
This  suggests that there would be  at most $\sim$ 260 sources/deg$^2$ Galactic sources
at this flux level, which corresponds to a 
$\sim 4 \times 10^{31}$ erg s$^{-1}$ source at 10 kpc assuming isotropic emission.
X-ray luminosity functions and spatial densities of 
quiescent dwarf novae have not been measured precisely ({\em 4, 6}), and
our results can place  an upper-limit for the quiescent dwarf nova population.
For example, a  combination of a $10^{-5}$ pc$^{-3}$ spatial density
and $10^{31}$ erg s$^{-1}$ average luminosity of quiescent dwarf novae,
which amounts to  $\sim 1000$ sources/deg$^2$ at $> 3\times10^{-15}$ ergs s$^{-1}$ cm$^{-2}$  ({\em 7}),  is
clearly ruled out.  To be consistent with our observation, either spatial density or average luminosity 
has to be smaller at least by a factor of four.

The paucity of Galactic point sources supports models of diffuse emission to explain
the Galactic ridge x-ray emission.
Then the next question is how to produce   plasmas 
with such a large energy density and high temperatures, and keep them
in the Galactic disk.
Theories  have been proposed to explain 
these problems in terms of  interstellar-magnetic reconnection ({\em 14}), 
interaction of energetic
cosmic-ray electrons ({\em 15})  or heavy-ions ({\em 16}) with interstellar
medium.  All of these models
require supernovae as  a mechanism of the energy input.
Incidentally, in the region around  $(l,b)\approx(28^\circ.55,-0^\circ.12)$, 
we found   an extended feature with a size of $\sim 2'$  
which is more conspicuous in the hard x-ray band  than in the soft band
(Fig.\  3).
This region corresponds to the southern end of 
an extended and  patchy radio source  named G28.60--0.13 ({\em   17}),
and the diffuse x-ray structure we found bridges three discrete radio 
patches  F, G and H ({\em 17}).
A high quality x-ray spectrum of the G28.60--0.13 
region by the ASCA satellite indicates that the
energy spectrum  is a single power-law without any iron line 
emissions ({\em 18}), which is reminiscence of the non-thermal acceleration taking place
in the supernova remnants such as SN1006 ({\em 19}) or  RX J1713.7-3946 ({\em 20}).
The   extended feature at $(l,b)\approx (28^\circ.55,-0^\circ.12)$, or 
G28.60--0.13 itself, may be  an aged supernova remnant.
Supernova remnants may be ubiquitous on the 
Galactic plane ({\em 17}) and these remnants may have an important 
role to generate the Galactic ridge hard  x-ray emission ({\em 21}).  

\newpage

 References and Notes
\begin{flushleft}
% 1. W. Forman {\em et al.}, {\em  Astrophys.\ J.\ Suppl.}    {\bf 38}, 357 (1978).\\
1. D. M. Worrall, F. E. Marshall, E. A. Boldt, J. H.  Swank, 
  {\em  Astrophys.\ J.} {\bf 255}, 111 (1982).\\
2. R. S. Warwick, M. J. L. Turner, M. G.  Watson, R.  Wilingale, 
 {\em  Nature} {\bf 317}, 218 (1985).\\
3. K. Koyama {\em et al.}, {\em  Publ. Astro. Soc. Japan} {\bf 38}, 121 (1986).\\
4. K. Mukai., K. Shiokawa,  {\em Astrophys. J.} {\bf 418}, 863 (1993).\\
5. R. Ottmann, J. H. M. M. Schmitt,  {\em Astron. \& Astrophys.} {\bf 256}, 421 (1992).\\
6. M. G. Watson,  in ``Annapolis Workshop on Magnetic Cataclysmic Variable'',
ASP Conference Series, vol. 157, 1999, p.291.\\
7. S. Yamauchi {\em et al.},  {\em Publ. Astro. Soc. Japan} {\bf 48}, L15 (1996).\\
8. H. Kaneda {\em et al.}, {\em Astrophys. J.} {\bf 491}, 638, (1997).\\
9. M. Sugizaki {\em et al.}, {\em Astrophys.\ J. Supp.} {\bf 134}, 77  (2001)\\
%10. R. F. Mushotzky, L. L.  Cowie, A. J.  Barger, A. J., K. A.  Arnaud, 
%  {\em  Nature} {\bf 404},  459 (2000).\\
10. R. Giacconi  {\em et al.}, {\em Astrophys.\ J.} {\bf 551}, 624 (2001)\\
11. Y. Ueda,  {\em et al.}, {\em Astrophys.\ J.}  {\bf 518},  656 (1999)\\
12. J. M. Dickey, F. J.  Lockman, {\em Ann. Rev. Astro. Astrophys.} {\bf 28}, 215 (1990).\\
13. T. M. Dame, D. Hartmann, P. Thaddeus, {\em Astrophys.\ J.}  {\bf 547}, 792 (2001)\\
14. S. Tanuma, {\em et al.}, {\em Publ. Astro. Soc. Japan} {\bf 51}, 161 (1999)\\
15. A. Valinia {\em et al.},  {\em Astrophys. J.} {\bf 543}, 733 (2000)\\ 
16. Y. Tanaka, T. Miyaji, G.  Hasinger, {\em Astronomische Nachrichten} {\bf 320}, 181
    (1999)\\
17. D. J. Helfand, T. Velusamy, R. H.   Becker, F. J. Lockman, 
 {\em Astrophys.\ J.} {\bf 341}, 151 (1989)\\
18. A. Bamba,  {\em et al.},  {\em  Publ. Astro. Soc. Japan} {\bf 53}, accepted (2001)\\
19. K. Koyama, {\em et al.},   {\em Nature} {\bf 378}, 255 (1995)\\
20. K. Koyama, {\em et al.},   {\em Publ. Astro. Soc. Japan} {\bf 49}, L7 (1997)\\
21. K. Koyama, S. Ikeuchi and K. Tomisaka,  {\em Publ. Astro. Soc. Japan} 
   {\bf 38}, 503 (1986)\\
22. We are grateful to  Drs. F. E. Marshall, K. Mukai,  and R. F. Mushotzky for
helpful comments which improved an earlier version of this Report.
We also thank Prof.\ Y. Tanaka and other anonymous referees for useful comments.
The observation was carried out under NASA's Chandra Guest Observer Program.
\end{flushleft}

\newpage 
% Figure 1
% Two images (all the field, central field)
\begin{figure}
%\centerline{
%\psfig{figure=plot_color_img2.ps,height=180mm,angle=0}
%}
\caption{
The {\sl Chandra\/} ACIS-I deep exposure image on the Galactic plane. 
Original image has been smoothed with an adaptive filter to make
both the point sources and diffuse emission clearly  visible. 
Color and contrast
are adjusted so that the point sources detected in the 0.5--3 keV band
and the  3 -- 8 keV band are conspicuous in red and blue,
respectively.
The hard x-ray diffuse emission is shown in blue.  
The image has 690 $\times$ 690 pixels with
the pixel size $1.''97$ square.
%Each
%dot corresponds to a single event, and photon energy is represented
%with different color.  Source search has been carried out in 0.5--8 keV, 0.5--3 keV, 
%and 3--8 keV independently, and all the 
%detected sources (at the 3 $\sigma$ confidence level) are encircled.
%Green and red ellipses indicate the sources detected in the soft band (106 sources) and hard 
%band (53 sources; 17 overlapped with soft sources) 
%respectively.  Black ellipses indicate the sources whose significances
%exceed 3 $\sigma$ only in the total band (0.5--8 keV; 33 sources).
%Total energy x-ray flux obtained from our observation was $\sim 7
%\times 10^{-11}$ erg cm$^{-2}$ s$^{-1}$ deg$^{-2}$ (2 -- 10 keV), which is 
%more or less consistent with previous values from ASCA observations.
%Total point source flux accounts for only $\sim 15 \%$ of that.
}
\end{figure}

% Figure 2   (log N - log S)
\begin{figure}
\centerline{\psfig{figure=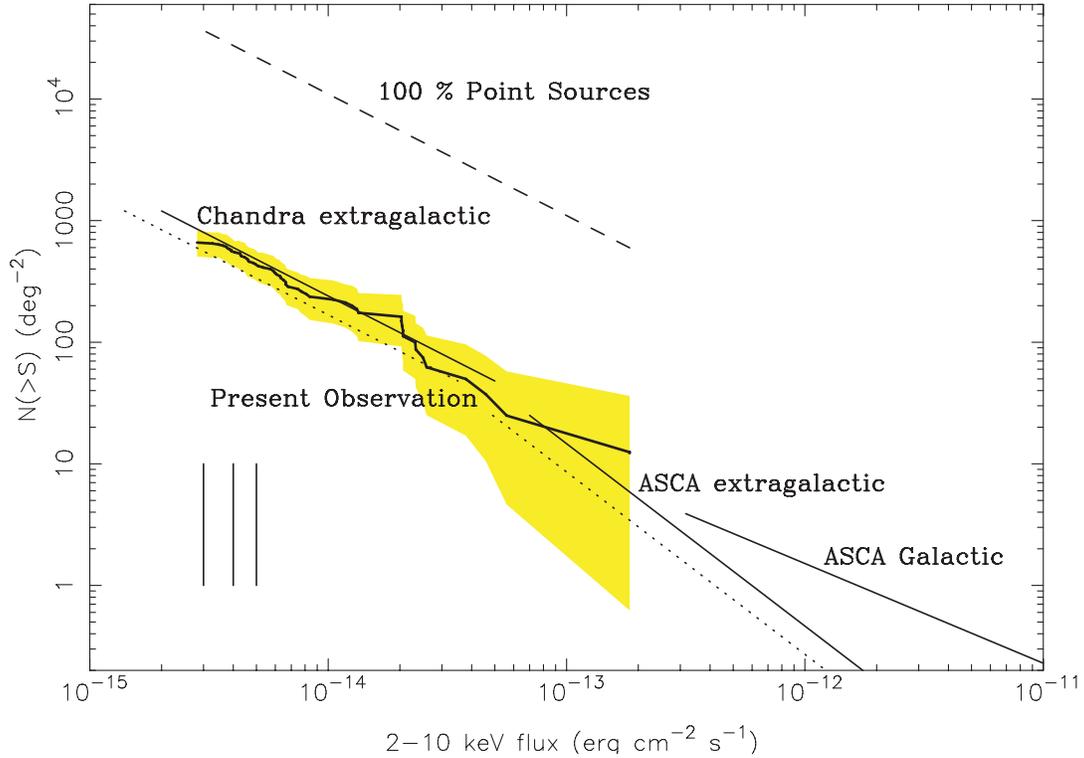,height=10cm,angle=270}}
\caption{
The number of point sources per unit area, $N$, brighter than
the threshold  flux in the 2 -- 10 keV band, $S$, is indicated as a function of $S$ (the $\log N - \log S$ curve).
The present Chandra $\log N - \log S$ curve 
on the Galactic plane is shown with the thick line, and the 90 \% error region (assuming
Poisson counting error) is indicated in yellow.
The upper dashed line shows the number of fictitious  point sources 
at a given flux $S$ that would have   accounted for  the total observed  x-ray energy flux
in the Chandra field of view ($\sim 1.1 \times 10^{-10} $
ergs s$^{-1}$ cm$^{-2}$ deg$^{-2}$ in 2 -- 10 keV).  
Three vertical lines at lower-left indicate the threshold energy fluxes
corresponding to 3, 4 and 4.5 $\sigma$ confidence of the source
detection.  For comparison,
the Galactic $\log N - \log S$ curve  by ASCA ({\em 9})
and extragalactic ones by Chandra ({\em 10}) and ASCA ({\em 11}) are shown.
The extragalactic $\log N - \log S$
curves  corrected for the effect of $\sim$ 30 \% flux reduction due to Galactic absorption with $N_H =
6 \times 10^{22}$ cm$^{-2}$ are  indicated with dotted line, which are consistent with
the present Galactic $\log N - \log S$ curve within 90 \% statistical error.
}
\end{figure}

%Figure 3
\begin{figure}
%\centerline{\psfig{figure=plot_diffuse_feature.ps,height=18cm,angle=0}}
\caption{Close-up of the diffuse feature at around  $(l,b) \approx
(28^\circ.55,-0^\circ.12)$.  
The 0.5 -- 3 keV image is represented in red and the 3 -- 8 keV image is in blue,
and both images are smoothed so as to enhance the diffuse feature.  Contour map
of the  3 -- 8 keV image is superimposed.  Point sources detected with
more than 3 $\sigma$ confidence either in the 0.5 -- 3 keV  band or the 3 -- 8 keV band
are indicated with crosses.
The image has 560 $\times$ 560 pixels with the pixel size 
 $0.''98$ square.
}
\end{figure}

\end{document}